\documentclass[12pt]{article}
\usepackage[]{epsfig}
\usepackage{amsfonts}
\usepackage{amsmath}
\usepackage{color}

\newcommand{\be}{\begin{equation}}
\newcommand{\ee}{\end{equation}}
\newcommand{\ba}{\begin{eqnarray}}
\newcommand{\ea}{\end{eqnarray}}

 \newcommand{\bea}{\begin{eqnarray}} \newcommand{\eea}{\end{eqnarray}}

\def\vr{{\check e}_r}
 \def\vteta{{\check e}_\theta}
 \def\vfi{{\check e}_\varphi}

\begin{document}
\title{\bf Holographic phase transition from dyons in an AdS black hole background}
\author{A.~R.~Lugo$^a$\thanks{Associated with CONICET}, E.~F.~Moreno$^b$  and
F.~A.~Schaposnik$^a$\thanks{Associated with CICBA}
\\
{\normalsize $^a\!$\it Departamento de F\'\i sica-IFLP, Universidad
Nacional de La Plata}\\ {\normalsize\it C.C. 67, 1900 La Plata,
Argentina}
\\
{\normalsize $^b\!$\it Department of Physics,West Virginia University}\\
{\normalsize\it Morgantown, West Virginia 26506-6315, U.S.A.} }
\date{\today}
\maketitle
%===================================================================
\begin{abstract}
We construct a dyon solution for a Yang-Mills-Higgs theory in a
4 dimensional Schwarzschild-anti-de Sitter black hole
background with temperature T. We then apply the AdS/CFT
correspondence to describe the strong coupling regime of a
$2+1$ quantum field theory  which undergoes a phase transition
exhibiting the condensation of a composite charge operator
below a critical temperature $T_c$.

\end{abstract}
%===================================================================
\section{Introduction}
It has recently been observed that the AdS/CFT correspondence
\cite{Maldacena}-\cite{Witten1998} provides a powerful tool to
study quantum critical dynamics in condensed matter models for
superconductivity,  superfluidity and Hall effect
\cite{Gubser1}-\cite{GubserP}. In this way, using the dual
classical gravity description, correlation functions in
strongly interacting systems can be calculated and the relevant
physical properties can be determined in a relatively simple
way, which has motivated an intense activity in the field (some
works related to the present approach are listed in \cite{F}).

Temperature is introduced in the gravity dual through the
presence of a Schwarzschild-AdS black hole, either  added as a
background or resulting from back reaction. Apart from gauge
fields, the models referred above include minimally coupled
scalar matter fields. In the Abelian case the ansatz for gauge
fields included an electrostatic potential $A_0$ which is
spatially independent at the boundary as also is the case for
the scalar. Spatially dependent droplet and vortex solutions
have been also studied in \cite{AJ}-\cite{MPS} and in the
non-Abelian case non-trivial gauge-field components  were
considered \cite{GubserNonAb},\cite{GubserP}. In general, such
classical solutions in the bulk exist below a certain critical
temperature leading to thermodynamically favored phases in the
the strongly  coupled quantum field theory on the boundary.

Following the ideas described above, we consider in the present
paper a model not yet studied: a dyon solution for
Yang-Mills-Higgs theory in a 4 dimensional
Schwarzschild-anti-de Sitter black-hole background metric. We
shall be guided by our previous work on monopole and dyon
solutions in AdS$_4$ space \cite{LS}-\cite{LMS} where, due to
the absence of singularities, the spherically symmetric
solutions were constructed in the whole radial domain
$\mathbb{R}^+$, while in the present case the radial variable
extend from the black hole horizon to infinity. The black hole
temperature, determined by the horizon, will become the
temperature of the field theory defined on the boundary. The
matter field  will lead to the non-Abelian gauge symmetry
breaking condensate while the field strength associated to the
surviving $U(1)$ symmetry will correspond to that of a dyon
with electric and magnetic charges.

The paper is organized as follows. In section 2 we present the
gravity-Yang-Mills-Higgs model and specialize to the case in
which the gravitational equations decouple from matter leading
to an AdS-Schwarzschild black hole background for the
gauge-matter system. Then in section 3 we analyze the
properties of the dyon solution considering both the cases of
an $S^2\times S^1$ and $\mathbb{R}^2 \times S^1$ boundaries. In
section 4 we proceed to describe the holography calculation of
relevant quantities in the dual field theory defined on the
boundary (taken as $\mathbb{R}^2 \times S^1$) and present our
numerical results. We end with a summary and discussion in
section 5.

\section{The gravity-Yang-Mills-Higgs system}

\subsection{The model}
We consider a gravity-Yang-Mills-Higgs system with gauge group
$SU(2)$ and the scalar field in the adjoint representation, in
a 4 dimensional space-time with Minkowski signature
$(-,+,+,+)$. The action takes the form
\be S = S_G + S_{YM} + S_H = \int d^4x\,\sqrt{|g|}\; ( L_G + L_{YM}
+ L_H ) \label{1} \ee
with \be L_G = \frac{1}{2\,\kappa^2}  \left(  R - 2\,\Lambda \right)
\label{2} \ee
\be L_{YM} = -\frac{1}{4e{}^2}\; F_{\mu\nu}^a F^{a\,\mu\nu}
\label{3} \ee
\be L_H = -\frac{1}{2}    D_\mu H^a \; D^\mu H^a - V(H) \label{4}
\ee
\be
 V(H) = \frac{\lambda}{4}\; ( H^a H^a - H_0{}^2 )^2
 \label{5}
 \ee
\be
\kappa^2\equiv 8\;\pi\; G_N \label{New}
\ee
Here   $G_N$ is the Newton constant, $e$ the gauge coupling
and $\Lambda$   the cosmological constant (with our conventions
$\Lambda < 0$ corresponds, in the absence of matter, to anti-de
Sitter space). The field strength   $F^a_{\mu \nu}$
($a=1,2,3$)  is defined as
\be
F^a_{\mu \nu} = \partial_\mu A_\nu^a - \partial_\nu A_\mu^a +
\varepsilon^{abc}A_\mu^b A_\nu^c \label{6}
\ee
and  the covariant derivative $D_\mu$ acting on the Higgs
triplet $H^a$ is given by
\be D_\mu H^a = \partial_\mu H^a + \varepsilon^{abc} A_\mu^b H^c
\label{7} \ee

The equations of motion that follow from (\ref{1}) are
\ba
E_{\mu\nu} + \Lambda\; g_{\mu\nu} &=&
\kappa^2 \; (T_{\mu\nu}^{YM} +  T_{\mu\nu}^H )\nonumber\\
{ D}_\rho D^\rho H^a &=& \frac{\delta V(H)}{\delta H^a}\nonumber\\
\frac{1}{e{}^2} D^\rho F^{a}_{\mu \rho} &=&
\varepsilon^{abc}\left( D_\mu H^b\right) H^c
\ea
where $E_{\mu\nu}$ is the Einstein tensor and the matter
energy-momentum tensor $T_{\mu\nu}$,
\be
T_{\mu\nu} = -2 \frac{\delta S}{\delta g^{\mu\nu}} \;\; ,
\ee
is given by
\ba
T_{\mu\nu}^{YM} &=& \frac{1}{e{}^2} F_{\mu \rho}^a
F^a_\nu{}^\rho
+ g_{\mu\nu}\; L_{YM}\nonumber\\
T_{\mu\nu}^H &=&   \; D_\mu H^a\; D_\nu H^a + g_{\mu\nu}\; L_H
\label{T} \ea

The most general static spherically symmetric form for the
metric in $3$ spatial  dimensions together with the
t'Hooft-Polyakov-Julia-Zee ansatz for the gauge and Higgs
fields in the usual vector notation reads \cite{LS}-\cite{LMS}
\ba g &=& - \mu(x)\; A(x)^2\; d^2 t + \mu(x)^{-1}\; d^2 r +
r^2\; d^2
\Omega_2\nonumber\\
\vec A &=& dt \; e\; H_0\; J(x)\; \vr - d\theta\; (1 - K(x) )\; \vfi
+ d\varphi \; (1 - K(x) )\; \sin\theta\; \vteta\nonumber\\
\vec H &=& H_0\; H(x)\; \vr
\label{ansatz}
\ea
where we  have introduced the dimensionless radial coordinate
$\; x\equiv e\; H_0\; r$ and we denote  the standard spherical
unit vectors as $\vr, \vfi,\vteta$. $H_0$ sets the mass scale
($[H_0] = m^1$).

Using this ansatz, the equations of motion take the form
\bea && \left( x\; \mu (x)\right)' = 1  + 3\; \gamma_0^2\;  x^2
-\kappa^2\, H_0{}^2 \left( \mu (x)\; V_1 + V_2 +
\frac{x^2}{2}\;\frac{J'(x)^2}{A(x)^2}
\right.\nonumber\\
&& \hspace{2.1 cm}  + \left. \frac{J(x)^2\, K(x)^2}{\mu(x)\,
A(x)^2} \right) \nonumber\\
&& x\; A'(x) = \kappa^2\, H_0{}^2 \left( V_1 + \frac{J(x)^2
K(x)^2}{\mu(x)^2 A(x)^2} \right) A(x) \nonumber \\
&&  \left( \mu (x)\; A(x)\; K'(x) \right)' = A(x)\; K(x) \left(
\frac{K(x)^2-1}{x^2} + H(x)^2
\right.\nonumber\\&& \hspace{3.9 cm} \left.
-\frac{J(x)^2}{\mu (x)\; A(x)^2}\right)\nonumber\\
&& \mu (x)\; A(x)\;\left( \frac{x^2\, J'(x)}{A(x)} \right)' = 2\; J(x)\; K(x)^2
\nonumber\\
&& \left( x^2\;\mu(x)\; A(x)\; H'(x) \right)' = A(x)\; H(x) \left( 2\; K(x)^2
+\frac{\lambda}{e{}^2}\; x^2\; ( H(x)^2 - 1) \right)\nonumber\\
\label{H}
\ea
where, for convenience, we have defined the dimensionless
parameter
\be \gamma_0^2 \equiv - \frac{\Lambda}{3 \,e{}^2\,
H_0{}^2} = \frac{1}{L^2 \,e{}^2\, H_0{}^2}\qquad,\qquad L^2
\equiv -\frac{3}{\Lambda} \label{gamacero} \ee and \ba
V_1 &=& K'(x)^2 + \frac{x^2}{2} H'(x)^2 \nonumber\\
V_2 &=& \frac{ (K(x)^2 -1)^2}{2\; x^2} +
\frac{\lambda}{4 e{}^2}\; x^2\; (H(x)^2 -1)^2
\ea

\subsection{The AdS-Schwarzschild black hole background}
In the $\kappa^2 \to 0$ limit the gravitational equations in
system (\ref{H}) decouple from the matter ones leading to a
solution of the metric $g$ of the form
\be g = - \mu_0(x)  d^2
t + \mu^{-1}_0(x)  d^2 r + r^2\; d^2 \Omega_2 \label{agu}
\ee
with
\be \mu_0(x) = 1 + \gamma_0^2 x^2 - \frac{R^3}{\gamma_0 x}
\label{background}
\ee
Here $R$ is a dimensionless parameter related to the black hole
mass. For $  R \ne 0$ the metric defined in (\ref{agu}) has an
horizon at $x= x_h$, i.e.,
\be \mu_0(x_h) = 0 \ee
with $\mu_0'(x_h) \ne 0$.  Written in terms of   the parameters
of the model $x_h$ reads
\be
\gamma_0 x_h =  \left(
\sqrt{\frac{R^6}{4} + \frac1{27}}  + \frac{R^3}{2}
\right)^{1/3}
-
\left(
\sqrt{\frac{R^6}{4} + \frac1{27}}  - \frac{R^3}{2}
\right)^{1/3}
\ee

In order to compute the black hole temperature one uses the
standard recipe \cite{HP}  leading to no conical singularity
at the horizon after Wick rotation to the Euclidean time
$\tau_E\equiv i\,t$. Hence, given (\ref{background}) and
imposing
\be \tau_E \sim \tau_E + \beta \ee
one identifies the black hole temperature as \be T \equiv
\frac1{\beta} = \frac{ | \mu_0'(x_h)|eH_0} {4\,\pi} =
\frac{1}{4 \pi L}\left(3 \gamma_0 x_h +\frac{1}{\gamma_0
x_h}\right)
\label{temp} \ee

\section{The dyon   in the black hole background}

\subsection{Properties of the solution}
From here on we shall consider, for simplicity, the BPS limit
of the potential (\ref{5}) which correspond to $\lambda/e^2 =
0$ with $H_0$ fixed. Taking the black hole metric
(\ref{background}) as a background    we are left with the
system
\begin{eqnarray}
&& \left( \mu_0 (x)\, K'(x) \right)' =  K(x)
\left( \frac{K(x)^2-1}{x^2} + H(x)^2
-\frac{J(x)^2}{\mu_0(x)}\right) \label{system1}\\
&&\mu_0 (x)\, \left(x^2\, J'(x) \right)' = 2\, J(x)\, K(x)^2
\label{system2}\\
&&\left( x^2\,\mu_0(x)\, H'(x) \right)' = 2 H(x)
    K(x)^2
\label{system3}
\end{eqnarray}

We shall look for a solution to (\ref{system1})-(\ref{system3})
regular at the horizon
\begin{equation}
K(x),\, H(x),\, J(x)/(x-x_h) \;\;\;\;\;
\text{regular at $x_h$}\label{origin}
\end{equation}
so that
\bea
H(x) &=&  h_0 + h_1 (x - x_h) +
{\cal O}[(x - x_h)^2]\nonumber\\
K(x) &=&  k_0 + k_1 (x - x_h) +
{\cal O}[(x - x_h)^2]\nonumber\\
J(x) &=&  j_h (x - x_h) +
{\cal O}[(x - x_h)^2]
\label{horj}
\eea
Concerning  $x \to \infty$,   the asymptotic behavior takes the form
\bea
|\vec H(x)| &=&     H_0\left( 1 {-}  \frac{H_1}{x^3} +
\frac{H_2}{x^5} +  \frac{H_3}{ x^{4 + 2\nu}} + \cdots \right) \label{Has}\\
 K(x) &=&
\frac{K_1}{x^{\nu +1}} + \frac{K_3}{x^{\nu +3}}\; , + \cdots
\;\;\;\; \nu \in\mathbb{R} \;\;\;\; \;\;\;\; \;\;\;\;  {\rm
as~x \to \infty} \label{K}\\
J(x) &=& J_0 + \frac{J_1}{x} + \ldots \label{asym}
\eea
Such behavior is consistent with the equation of motion
(\ref{system1}) provided that the following condition holds
\be \frac1{\gamma_0^2} =   \nu(\nu +1) \label{eq} \ee

Equation (\ref{eq}) has two solutions,
\be
\nu_\pm = -\frac12 \pm \frac12\sqrt{1 + \frac{4}{\gamma_0^2}}
\ee
Only the $\nu_+$ root gives an acceptable asymptotic behavior
for $K(x)$ so that
\be K(x)  \sim \frac{K_1 }{x^{\nu_+ +1}}  = \frac{K_1
}{x^{\frac12\left(1 +  \sqrt{1
+\frac{4}{\gamma_0^2}}\,\right)}} \;\;\;\; \;\;\;\; \;\;\;\;
{\rm as}~x \to \infty \ee
Being the $SU(2)$ gauge symmetry of the Lagrangian
spontaneously broken by the expectation value of the scalar
field, we shall define the field strength associated with the
surviving $U(1)$ symmetry as in \cite{Polyakov}
\be {\cal F}^{U(1)}_{\mu\nu} \equiv
\frac{H^a}{H_0} F_{\mu\nu}^a \ee
From this expression one finds the U(1) magnetic and electric
fields
\be B^i = \frac12 \frac{\varepsilon^{ijk}}{\sqrt{
g^{(3)}}}{\cal F}^{U(1)}_{jk} \; , \;\;\; E_i = {\cal
F}^{U(1)}_{i0} \label{B} \ee
and the magnetic and electric charges
\bea Q_m &=& \int_V  d^3 x \sqrt{ g^{(3)}} \nabla^{(3)}_i
B^i
\label{Qm}\\
Q_e &=& \int_V d^3 x  \sqrt{ g^{(3)}} \nabla^{(3)}_i E^i
\label{Qe}
\eea
where $\nabla^{(3)}_i$ is the 3-dimensional covariant
derivative. In order to compute $Q_m$ one just needs the
asymptotic value of ${\cal F}^{U(1)}_{ij}$ since
 eqs.(\ref{B})-(\ref{Qm}) imply
\be
Q_m =  \frac12  \int_{\partial V=S^2} {\varepsilon^{ijk}} {\cal F}^{U(1)}_{jk} dS_i
\label{Qm2}
\ee
Now, from the asymptotic conditions (\ref{asym}) one has
\be
{\varepsilon^{ijk}} {\cal F}^{U(1)}_{jk}  \sim
\frac{1}{x^2}
\vr \delta^{i{r}} \;\;\;\;\;\;\;\;\;\;
 {\rm for~} {x \to \infty}
\ee
and then
\be Q_m = -\frac1e\int_{S^2}   d\Omega = -\frac{4\pi}{e}
\label{Q} \ee
As expected, the spherically symmetric solution has a quantized
$n=-1$ magnetic charge (in units of $4\pi/e$).

Concerning the electric charge, one can see from the asymptotic
behavior of $J$, eq.(\ref{asym}), that $J_0$ sets the charge
scale and  $J_1$  determines its value. One can write
\be Q_e = \frac{1}{e} {\cal Q} (e^2/\lambda, J_0) =- \frac{4
\pi}{e} J_1\ee
with ${\cal Q}$ a dimensionless function to be determined
numerically. As for the Julia-Zee flat space dyon, the electric
charge is not quantized at the classical level. As we shall
see, $J_0$ and $J_1$ will be identified, through holography,
with the chemical potential and charge density of the strongly
coupled system respectively.

Let us now find the expression for the dyon  action which will
be connected with the free energy of the field theory defined
on the boundary. Inserting ansatz (\ref{ansatz}) in
eqs.(\ref{3})-(\ref{4}) one obtains for the Yang-Mills and
Higgs Lagrangians
\bea L_{YM} &=& -\frac{e^2H_0^4}{2} \left( -\frac{{J'}^2}{A^2} - 2
\frac{J^2K^2}{x^2\mu A^2} + 2\frac{\mu {K'}^2}{x^2} + \frac{(K^2 -
1)^2}{x^4} \right)
\nonumber\\
L_{H} &=& -\frac{e^2H_0^4}{2}\left( \mu {H'}^2 +
2\frac{K^2H^2}{x^2}\right)
\eea
Then, by analytic continuation, the Euclidean action takes the
form
\be
S_E = - \int_0^\beta \!\!d\tau_E\int_V d^3x \sqrt g \left(  L_{YM} + L_H\right)
\label{given}
\ee

\subsection{From $\mathbf{S^2}$ to $\mathbb{R}^2$}

As stated in the introduction, our aim  is to calculate the
thermodynamics  of the strongly interacting field theory
defined on the boundary ${\partial M}$ using the dual classical
description of the gravity system governed by the action
(\ref{1}) in the 4 dimensional manifold $ {M}$; now, according
to (\ref{agu}), ${\partial M} = {S^2} \times {S^1}$. From the
condensed matter viewpoint, it is more relevant to study the
finite temperature field theory defined on the plane
${\mathbb{R}}^2$ and not on the sphere $S^2$. To this end, one
can follow the approach  in \cite{Witten2} and proceed  to a
change coordinates $(t,x,\theta,\varphi)\rightarrow (\tau, y ,
x^1,x^2)$ or, equivalently, $(t,x,\theta,\varphi)\rightarrow
(\tau,y,\rho,\varphi)$,
\bea \tau &=& R\; t\cr y &=&
\frac{\gamma_0}{R}\;x\cr x^1 &=& 2\;R\;\tan \frac{\theta}{2}\;
\cos\varphi = \rho\; \cos\varphi\cr x^2 &=& 2\;R\;\tan
\frac{\theta}{2}\;\sin\varphi = \rho\; \sin\varphi \eea
together with the field redefinitions
\bea \tilde A(y) &=&
A(x)\cr \tilde\xi(y) &=& R^{-2}\;\xi(x)\cr \tilde H(y) &=&
H(x)\cr \tilde K(y) &=& R^{-1}\;K(x)\cr \tilde J(y) &=&
R^{-1}\;J(x) \eea
where $R$ was introduced in (\ref{background}). After this
change,   ansatz (\ref{ansatz}) becomes
\bea g &=& - \tilde f_R(y)\; \tilde A(y)^2\; d^2 \tau +
L{}^2\;\frac{ d^2 y}{\tilde f_R (y)} + L^2\; y^2 \frac{d^2\vec
x}{(1+\frac{\rho^2}{4\,R^2})^2}\cr \vec A &=& d\tau \; e\;
H_0\; \tilde J(y)\; \left(\frac{4\,R\,\rho}{\rho^2 +
4\,R^2}\;\check e_\rho + \frac{4\,R^2 - \rho^2}{\rho^2 +
4\,R^2}\;\check e_3\right)\cr &+& \frac{4\,R^2}{\rho^2 +
4\,R^2}\;\left(\frac{1}{R} -\tilde K(y)\right)\; \left[
\left(\frac{4\,R^2 - \rho^2}{\rho^2 + 4\,R^2}\;\check e_\rho -
 \frac{4\,R\,\rho}{\rho^2 + 4\,R^2}\;\check e_3\right)\;
\rho\;d\varphi \right.\cr &-& \left. \vphantom{\frac1R}
\check e_\varphi\;d\rho\right]\cr
\vec H &=& H_0\; \tilde H(y)\; \left(\frac{4\,R\,\rho}
{\rho^2 + 4\,R^2}\;\check e_\rho +
\frac{4\,R^2 - \rho^2}{\rho^2 + 4\,R^2}\;\check e_3\right)
\label{ansatz2}
\eea
where
\be \tilde f_R(y) = f_R(y) + \tilde\xi(y)\qquad,\qquad
f_R(y)\equiv \frac{1}{R^2} + y^2 - \frac{1}{y} \ee
and the equations of motion for the Einstein-Yang Mills-Higgs
system, eqs.(\ref{H}) take the form,
\bea && \left( y\; \tilde\xi
(y)\right)' = -\kappa^2\,H_0{}^2\; \left( \tilde f_R (y)\;
\tilde V_1 + \tilde V_2 + \frac{y^2}{2}\;\frac{\tilde
J'(y)^2}{\tilde A(y)^2} + \frac{\tilde J(y)^2\,\tilde
K(y)^2}{\tilde f_R(y)\,\tilde A(y)^2}\right)\cr && y\; \tilde
A'(y)  = \kappa^2\, H_0{}^2 \left( \tilde V_1 + \frac{\tilde
J(y)^2\,\tilde K(y)^2}{\tilde f_R(y)^2\,\tilde A(y)^2}
\right)\;\tilde A(y) \label{eqgra2}
\\
 && \left( \tilde f_R(y)\, \tilde A(y)\,\tilde K'(y) \right)'
 =
\tilde A(y)\; \tilde K(y)\; \left( \frac{\tilde K(y)^2 -\frac{1}{R^2}}{y^2} +
 \frac{1}{\gamma_0{}^2}\;\tilde H(y)^2 \right.
\cr
&& \hspace{4.3 cm} \left.-\frac{1}{\gamma_0{}^2}\;\frac{\tilde J(y)^2}{\tilde f(y)\,\tilde A(y)^2} \right)
\cr
&& \left( y^2\,\tilde f_R(y)\,\tilde A(y)\,\tilde H'(y) \right)'= 2\tilde A(y)\;\tilde H(y)\;
  \tilde K(y)^2    \cr
&& \tilde f_R(y)\,\tilde A(y)\;\left( \frac{y^2\tilde J'(y)}{\tilde A(y)} \right)' =
2\;\tilde J(y)\,\tilde K(y)^2\label{eqmatt2}
\eea
where
\bea\label{poten2}
\tilde V_1 &=& \gamma_0{}^2\;\tilde K'(y)^2 + \frac{y^2}{2}\;\tilde H'(y)^2 \cr
\tilde V_2 &=& \gamma_0{}^2\;\frac{ (\tilde K(y)^2 -\frac{1}{R^2})^2}{2\; y^2}
\eea

As explained above, our aim is to consider the case in which
the gravitational equations decouple from the matter ($\kappa^2
\to 0$) leading to a background metric of the form (\ref{agu}).
The relevant  equations of motion for the gauge and Higgs field
in the $R \to \infty$ limit are then
\begin{align}
&\left( \tilde f_\infty(y)\,\tilde K'(y) \right)' = \tilde K(y)\, \left(
\frac{\tilde K(y)^2}{y^2} + \frac{1}{\gamma_0{}^2}\;\tilde H(y)^2
-\frac{1}{\gamma_0{}^2}\;\frac{\tilde J(y)^2}{\tilde f_\infty(y)} \right)
\label{eq1}\\\
&\left( y^2\,\tilde f_\infty(y)\,\tilde H'(y) \right)' = 2\,\tilde H(y)\,\tilde K(y)^2 \\
&\tilde f_\infty(y)\,\left( y^2\tilde J'(y)\right)' = 2\;\tilde
J(y)\,\tilde K(y)^2 \label{eqmatt22}
\end{align}
\[
\tilde f_\infty(y) = y^2 -\frac{1}{y}
\]

Let us end this section by noting that the black hole
temperature resulting from rescaling (\ref{temp})  and taking
the $R \to \infty$ limit  takes the form
\be \hat T \equiv \frac{T}{e\, H_0}=\frac{3 \gamma_0}{4\pi}
\label{rrr} \ee

\section{Holographic correspondence}

According to the AdS/CFT correspondence
\cite{Maldacena}-\cite{Witten1998}, properties of the dual 3
dimensional field theory defined on the boundary can be read
from the  behavior of the solution of the system in the bulk.
In this approach temperature $\hat T$ in (\ref{rrr})
corresponds to the temperature  of the $d=3$ system.

Let us first consider the vacuum expectation value in the $d=3$
field theory for the scalar operator ${\cal O}_H$, dual to the
field $H$ defined on the bulk. It follows from the
identification $ \langle O_H \rangle \sim  H_1 $  with $H_1$
defined in eq.(\ref{Has}).

The asymptotic behavior of the $A_0$ component of the gauge
field allows one to identify   the chemical potential $\mu$ and
the charge density $\rho$ of the 3 dimensional system. Indeed,
given the ansatz for the $A_0$ component of the gauge field,
eq.(\ref{ansatz}), the $U(1)$ electric field as defined  in
(\ref{B}) reads
\be E_i = {\cal F}^{U(1)}_{i0} = \frac{H^a(x)}{H_0} F^{a}_{i 0}
 = eH_0 H(x)\partial_i  J(x)
\ee
(the $\vec H\cdot \vec A_0 \wedge \vec A_i$ contribution
coming from the field strength commutator vanishes because
$\vec H \parallel   \vec A_0 $).

Since $\lim_{x\to \infty} H(x) = 1$, the electric field
at infinity is given by
\be
\lim_{x \to \infty} E_i   =
eH_0\lim_{x \to \infty} \partial_i J(x)
\ee
We have now to expand $J(x)$ for large $x$ and look for the
$1/x$ term which is the one leading to a nontrivial electric
flux (i.e., leading to the dyon charge). The corresponding
coefficient defines the charge density:
\be \rho = -  L H_0 \tilde J_1 \label{71} \ee
Here and in what follow we use the coefficients of the
asymptotic expansions defined in ${\mathbb{R}^2}\times S^1$.
Following \cite{GP} we shall define a normal component $\rho_n$
by analyzing the  expansion for $\tilde J(y)$ at the horizon
using eq.(\ref{horj}),
\be \rho_n =
\tilde j_h \label{72} \ee
and a superconducting charge density in the form
\be \rho_s = \rho - \rho_n \label{73} \ee

The constant value $\tilde J_0$ that $A_\tau$ takes at the
boundary at infinity   is related to the chemical potential
through the formula
\be \mu = H_0 \tilde J_0 \ee

\section{Numerical Analysis}
We shall restrict the numerical calculations to the case in
which the boundary is $S^1 \times{\mathbb{R}}^2$. We shall
solve system (\ref{eq1})-(\ref{eqmatt22})   using the
relaxation method \cite{relaxation} which determines the
solution starting from an initial guess and improving it
iteratively. The natural initial guess is the
Prasad-Sommerfield solution \cite{PS} in flat space. Following
this procedure, we have found regular solutions for different
values of $\gamma_0$ i.e. of the temperature $\hat T$.

Figures 1 and 2 show a representative dyon solution that exists
starting at the horizon. One can see that the scalar rapidly
attains its symmetry breaking constant value while the magnetic
and electric fields (associated to $K$ and $J$ respectively)
concentrate in a spherical shell starting at the horizon.
Figure 3 shows a remarkable property of the coupled system of
non linear differential equations, namely the existence of
several solutions satisfying the   appropriate boundary
conditions, each one corresponding to a different value of
charge density $\rho$.  This phenomenon was already encountered
in \cite{AJ} for the case of a $U(1)$ gauge field coupled to a
complex scalar in an AdS$_4$-Schwarzschild background. As it
also happens in the $U(1)$ case, we have found that for
increasing values of $\rho$ the solutions are distinguished by
an increasing number of nodes $n$. Now, since evaluation of the
free energy shows that it increases with the number of nodes
(see below), we conclude that $n>1$ solutions are
thermodynamically unfavorable and therefore we shall
concentrate hecenforth on the zero node solution.

\vspace{1 cm}

{\epsfxsize=5.2 in \epsffile{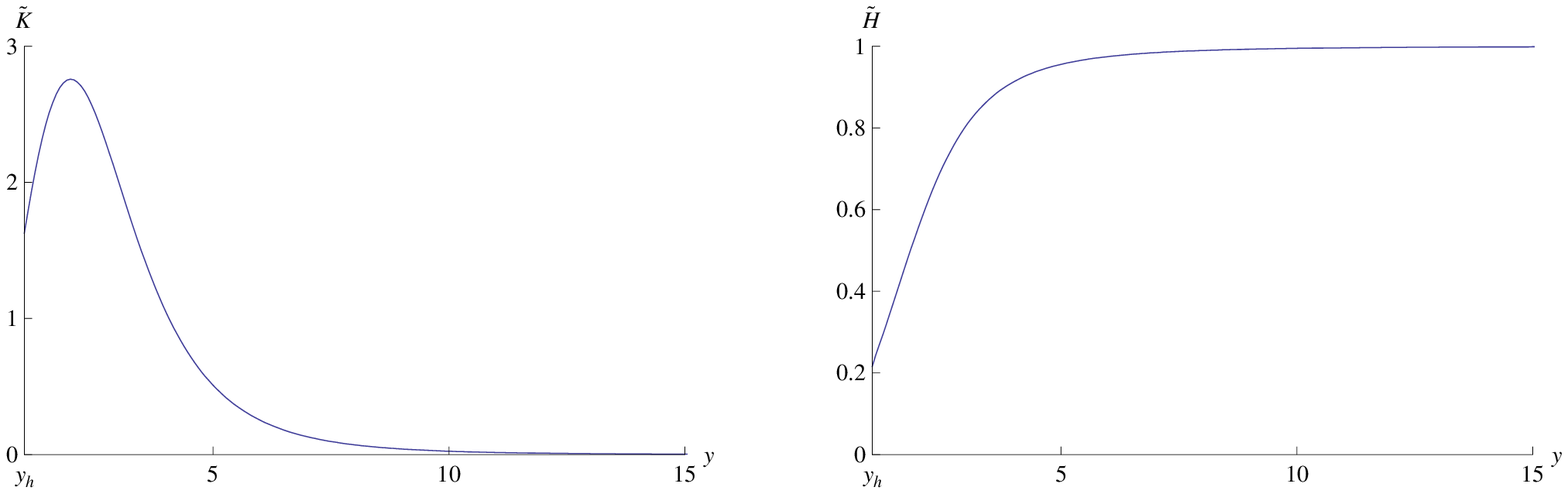}}

\noindent Figure 1: The solution for  $\tilde K$ and $\tilde H$
as a function of $y $ or $\tilde J_0=6$ and $\gamma_0=0.2$
($\hat T = 3\gamma_0/(4 \pi)$). The solution exists starting at
the horizon $y=y_h=1$.

\vspace{1.5 cm}

\centerline{\epsfxsize=2.5  in \epsffile{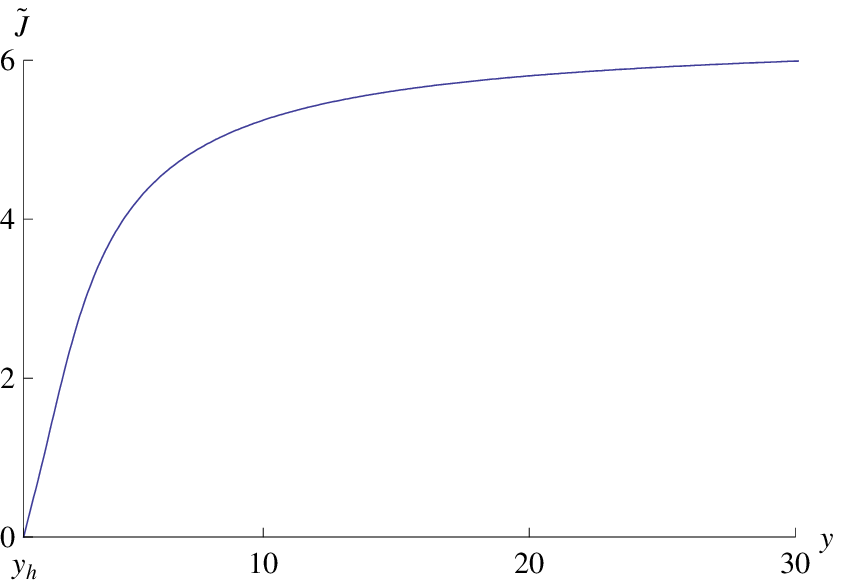}}

\noindent Figure 2: The solution for  $\tilde J$ as a function
of $y$ for $\tilde J_0=6$ and $\gamma_0=0.2$ ($\hat T =
3\gamma_0/(4 \pi)$).

\vspace{1. cm}

\centerline{\epsfxsize=3.9  in \epsffile{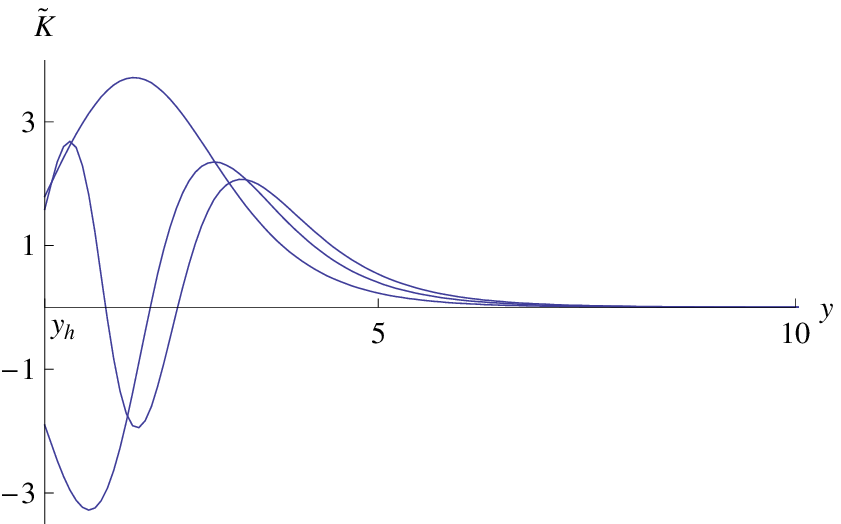}}

\noindent Figure 3: Different  solutions for  $\tilde K$
corresponding to different values of the charge density $\rho$.
For increasing values of $\rho$ the solutions are distinguished
by the increasing number of nodes.

\vspace{0.5 cm}

The behavior of the condensate is shown in figure 4 where the
asymptotic coefficient $H_1 \sim \langle O_H \rangle$ in the
asymptotic expansion (\ref{Has}) of the Higgs field is plotted
as a function of the temperature ({normalized in units of
$eH_0$}). In figures 5 and 6 we plot the coefficient $\tilde
K_1$ in the asymptotic expansion of function $\tilde K$,  while
figure 7 shows the fraction $\rho_s/\rho$, as defined in
eqs.(\ref{71})-(\ref{73}), related to the behavior of the
electric field at the horizon and at infinity. From these
figures we conclude that a finite temperature continuous
symmetry breaking transition takes place so that the system
condenses below a critical temperature $\hat T_c$. By fitting
the curves we see that near $T_c$ one has a typical second
order phase transition with power behavior of the form
\begin{eqnarray}
\tilde H_1 &\propto& (\hat T_c - \hat T)  \nonumber\\
\tilde K_1   &\propto& (\hat T_c - \hat T)^{1/2}
\; \;\;\;\; {\rm as~} \hat T \to \hat T_c \nonumber \\
\frac{\rho_s}{\rho}  &\propto& (\hat T_c - \hat T)
\end{eqnarray}
Notice that $\tilde K_1$ has a critical exponent 1/2 typical of
second order transitions.

One can see in figure 4 that at low temperature $\hat H_1$
appears to diverge. However, one should take into account that
when the condensate becomes very large, the backreaction on the
metric can no longer be neglected as we did by choosing a fixed
Schwarzschild-AdS background metric.

\centerline{\epsfxsize=4 in \epsffile{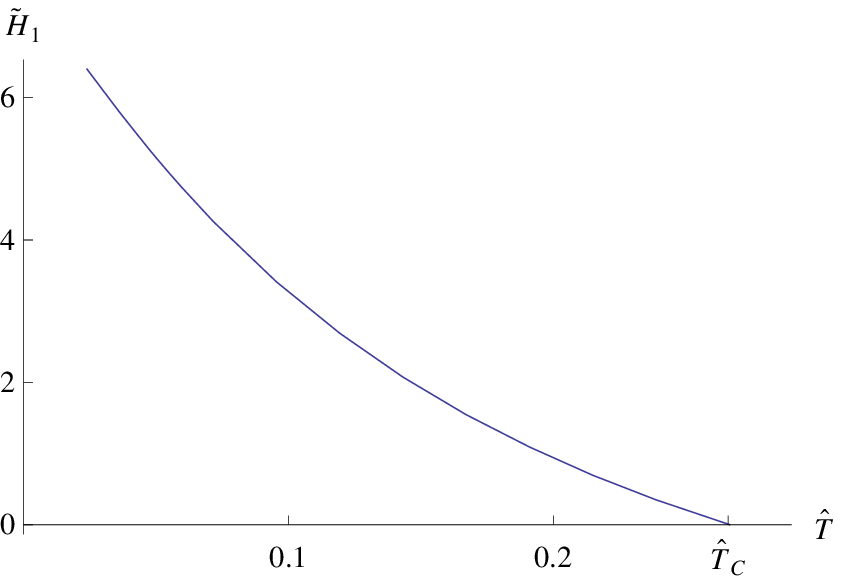}}

\noindent Figure 4: The $\tilde H_1$ coefficient in the Higgs
field asymptotic expansion as a function of the temperature. We
have taken $\tilde J_0=6$ and for this value the critical
temperature is $\hat T_c = 0.26649...$.

~

\vspace{0.5 cm}

\centerline{\epsfxsize=4 in \epsffile{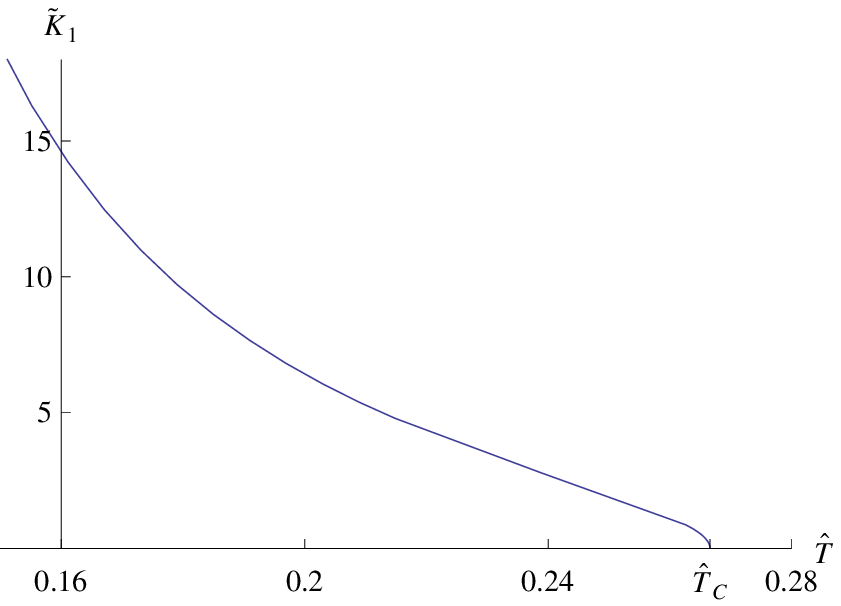}}

\noindent Figure 5: The $\tilde K_1$ coefficient in the gauge
field asymptotic expansion as a function of the temperature for
$\tilde J_0=6$.

\vspace{0.5 cm}

\centerline {\epsfxsize=4. in \epsffile{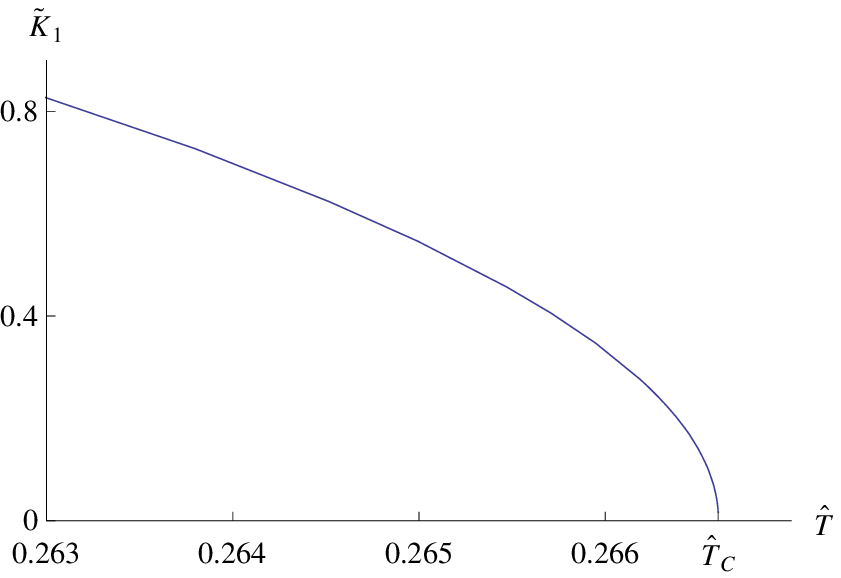}}

\noindent Figure 6: A zoomed view of $\tilde K_1$ near the
critical point. One can see that its behavior near $T = T_c$
corresponds to $\sqrt{\hat T_c-\hat T}$.

\centerline{\epsfxsize=4 in \epsffile{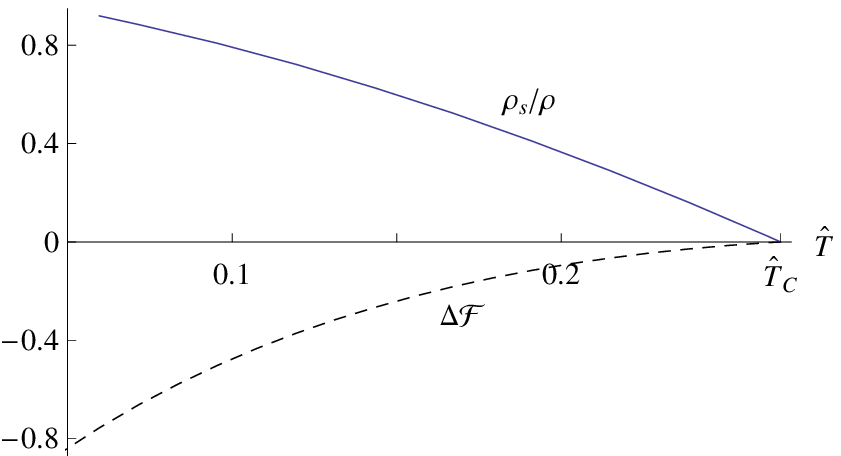}}

\noindent Figure 7: The solid line represents the fraction
$\rho_s/\rho$ of the superconducting and the total charge
densities. The dashed line the free energy difference between
the condensed and the uncondensed phases. At $\hat T= \hat
T_c$, $\rho_s/\rho$ vanishes linearly. \vspace{0.5 cm}

The free energy ${\cal F}$  is given by the on-shell action,
\be
{\cal F} = \left. T S_E\right|_{on~shell}
\ee
Using the Euclidean action as given in eq.(\ref{given}) we have
numerically computed   ${\cal F}$ for the dyon solution and
compared it with the free energy in the uncondensed phase which
corresponds to the solution $ \tilde H(y) = 1, \tilde K(y) =0,
\tilde J(y) = \tilde J_0(1 - 1/y) $. We plot in figure 7 the
free energy density difference  $\Delta{\cal F}$, normalized in
units  of $H_0^2/(2 L)$, which can be seen to be  continuous at
$\hat T=\hat T_c$.

\vspace{0.5 cm}

\section{Summary and discussion}
We have applied the AdS/CFT correspondence to describe the
strong coupling regime of a $2+1$ quantum field theory  which
undergoes a phase transition exhibiting the condensation of a
composite charge operator below a critical temperature $T_c$.
The dual gravity theory consists of an $SU(2)$ Yang-Mills-Higgs
model in a Schwarzschild-AdS$_4$ black hole with temperature
$T$.

Breaking the gauge symmetry down to $U(1)$ we started by
constructing a spherically symmetric soliton solution in the
bulk, having both magnetic and electric charge. By choosing
appropriate conditions at the horizon and at infinity, we found
a family of dyon solutions leading to the existence of a
condensate, below a critical temperature,  in the dual field
theory on the boundary which, after an appropriate scaling,
becomes $\mathbb{R}^2\times S^1$.

By calculating the free energy ${\cal F}$ we have seen that the
condensed phase is thermodynamically favored and we were able
to determine which one of the dyon family solutions had the
lowest ${\cal F}$-value. We have also calculated the power
behavior of the condensate, typical of a second order phase
transition.

There are many issues that we have not discussed in our work
and that deserve a thorough study. To begin with, one should
consider the applicability of our results to describe the
behavior of an actual condensed matter system. Also, since we
have found on the gravity side classical solutions with
quantized magnetic flux, we expect that the issue of
confinement through the appearence of chromoelectric flux tubes
in the dual strongly coupled QFT could take place. In this
respect, it should be   of interest to consider monopole
solutions in Schwarzschild-AdS$_5$ black hole backgrounds so
that the dual theory is defined in three spatial dimensions. We
hope to discuss theses issues in a future publication.

\vspace{1 cm}

\noindent\underline{Acknowledgments} We would like to thank
Jorge Russo for discussions and helpful suggestions and Tameem
Albash for a clarifying comment. A.R.L. is grateful with Borut
Bajc for helpful discussions. This work was partially supported
by PIP6160-CONICET, BID 1728OC/AR PICT20204-ANPCYT grants and
by CIC and UNLP, Argentina.

\end{document}